\documentclass[11pt, leqno]{article}
\newtheorem{prop}{Proposition}
\newtheorem{lem}{Lemma}

\newtheorem{cor}{Corollary}
\newtheorem{thm}{Theorem}
\newcommand{\be}{\begin{equation}}
\newcommand{\ee}{\end{equation}}
\newcommand{\ba}{\begin{array}}
\newcommand{\ea}{\end{array}}

\newcommand{\bea}{\begin{eqnarray*}}
\newcommand{\eea}{\end{eqnarray*}}
\newcommand{\bean}{\begin{eqnarray}}
\newcommand{\eean}{\end{eqnarray}}
\newcommand{\proof}{\vspace{1ex}\noindent{\em Proof}. \ }
\def\ds{\displaystyle}
\def\nm{\noalign{\medskip}}

\newcommand{\R}{\mbox{\bf R}}
\def\Box{\leavevmode\vbox{\hrule
     \hbox{\vrule\kern5pt\vbox{\kern5pt}%
           \vrule}\hrule}}
\newcommand{\square}{\hfill$\Box$}
\begin{document}

\title{Identifying of the refractive index for the acoustic equation at
fixed frequency}

\author{Christian DAVEAU   \and Abdessatar KHELIFI \and Anton SUSHCHENKO}
\maketitle \abstract{ In this paper we determine a formula for
calculating the refractive index ${\bf n}$ for the acoustic
equation from the partial Dirichlet to Neumann map(DN) associated
to ${\bf n}$. We apply these results to identify locations and
values of small volume perturbations of this refractive index at
fixed frequency $\omega$. }

\begin{center}
 {\bf Identification de l'indice de r\'efraction de l'\'equation
acoustique \`a fr\'equence
 fixe}
\end{center}

\begin{center}
{\bf R\'esum\'e}
\end{center}
Dans cette Note nous d\'eterminons une formule pour calculer
l'indice de r\'efraction ${\bf n}$ pour l'\'equation acoustique
\`a partir de l'application Dirichlet-to-Neumann partielle (DN)
associ\'ee \`a ${\bf n}$. Nous nous appliquerons ces r\'esultats
pour identifier les locations et les valeurs des petites
perturbations volumiques associ\'ees \`a cet indice de
r\'efraction pour une fr\'equence $\omega$ fixe.

\section*{Version fran\c{c}aise abr\'eg\'ee}
Cette Note traite un probl\`eme inverse pour l'\'equation
acoustique avec fr\'equence fixe. Le but est d'identifier l'indice
de r\'efraction associ\'e apr\`es avoir d\'eriver une formule
appropri\'ee \`a l'aide de l'application Dirichlet-to-Neumann
partielle (DN) associ\'ee \`a ${\bf n}$ et de donner plus
explicitement cette identification. Concernant l'utilisation de
l'application Dirichlet-to-Neumann partielle en probl\`eme
d'identification, nous pouvons trouver, par exemple, le travail De
Kohn et Vogelius ~\cite{KV}.\\

Soit $\Omega \subset \R^3 $ un domaine born\'e avec un bord de
classe $ C^2 $. Nous noterons $ \nu $ la normale unitaire sortante
du bord $ \partial\Omega $. Soit $ \Gamma $ une partie ouverte
r\'eguli\`ere de la fronti\`ere $ \partial\Omega $. Supposons que
$ \Omega $ contient un nombre fini d'inhomog\'en\'et\'es, chacune
de la forme $ z_j + \alpha B_j $, o\`u $ B_j \subset \R^3 $ born\'ee et contenant l'origine.\\
Soit $ {\bf n}(x) \in C^0(\Omega) $ l'indice de r\'efraction non
perturb\'e. Nous supposons que $ {\bf n}(x) $ est connue sur un
voisinage de $\partial\Omega $. Notons par $ {\bf n}_j(x) \in
C^0(\overline {z_j + \alpha B_j}) $ l'indice de r\'efraction de la
j-\`eme inhomo\'e'et\'e, $z_j + \alpha B_j$.\\
Consid\'erons l'\'equation acoustique, \`a fr\'equence fixe, en
pr\'esence d'inhomog\'en\'e\'et\'es
 \[
 \begin{array}{l}(\Delta + \omega^2 {\bf n}_\alpha) u_\alpha = 0 \mbox{ et } \Omega \\\nm u_\alpha|_{\partial \Omega} =
f \in \widetilde{H}^{\frac{1}{2}}(\Gamma), \end {array}
\]

et d\'efinissons l'application Dirichlet-to-Neumann partielle
associ\'ee \`a $ {\bf n}_\alpha $ par: $ \ds \Lambda_{{\bf
n}_\alpha}(f) = \frac{\partial u_\alpha}{\partial\nu}|_{\Gamma} $
pour tous $ f \in \widetilde{H}^{\frac{1}{2}}(\Gamma)$. Ici
$\widetilde{H}^{\frac{1}{2}}(\Gamma)$
d\'esigne l'espace trace. \\

Bukhgeim et Uhlmann \cite{BU} ont utilis\'e les solutions
complexes de l'optique g\'eom\'etrique pour prouver que la
connaissance des donn\'ees partielles de Cauchy pour l'\'equation
de  Schr\"{o}dinger d\'etermine le potentiel de mani\`ere unique
sachant que  le potentiel $ q \in L^{\infty}(\Omega) $ est connue
dans un voisinage du bord. Nous pouvons utiliser leurs travaux
pour montrer notre version physique et nous utilisons ainsi ces
solutions complexes de l'optique g\'eom\'etrique construites dans
leur travaux pour prouver notre
m\'ethode de reconstruction.\\

Ensuite, nous d\'erivons une formule de calcul de l'indice de
r\'efraction $ {\bf n} $ \`a partir de l'application
Dirichlet-to-Neumann partielle associ\'e \`a $ {\bf n} $ et $
\Gamma $ en modifiant la proc\'edure de reconstruction de
Nachman~\cite{Nach}. Enfin, nous profitons des propri\'et\'es des
noyaux des op\'erateurs ainsi introduites pour pr\'esenter plus
explicitement cette formule de reconstruction. Nous consid\'erons
le probl\`eme inverse d'identification des lieux de petites
perturbations volumique de l'indice de r\'efraction et nous
appliquons nos r\'esultats pour r\'eduire ce probl\`eme inverse au
calcul de transformation de Fourier inverse.

\section{Problem formulation}
 The paper deals with an inverse problem for
 the acoustic equation with fixed frequency. The object is to identify the associated refractive
 index after deriving a suitable formula by using the partial Dirichlet to Neumann map(DN) associated
to ${\bf n}$ and to give more explicitly this identification.
Concerning the use of the partial Dirichlet to Neumann map in
recovering problem we can find, for example, the work
of Kohn and Vogelius ~\cite{KV}.\\
Let $\Omega \subset \R^3$ be a bounded domain with $C^2$ boundary.
We denote by $\nu$ the unit-outer normal to $\partial \Omega$. Let
$\Gamma$ be a smooth open subset of the boundary $\partial \Omega$
and $\Gamma_c$ denotes $\partial \Omega \setminus
\overline{\Gamma}$. We suppose throughout that
$\R^3\backslash\overline{\Omega}$. Introduce the trace space
$$\widetilde{H}^{\frac{1}{2}}(\Gamma) = \Bigr\{ f \in
{H}^{\frac{1}{2}}(\partial \Omega), f \equiv 0 \mbox{ on }
\Gamma_c \Bigr\}.$$ Here and in the sequel we identify $f$ defined
only on $\Gamma$ with its extension by $0$ to all $\partial
\Omega$. It is known that the dual of
$\widetilde{H}^{\frac{1}{2}}(\Gamma)$ is
${H}^{-\frac{1}{2}}(\Gamma)$.\\ Assume that $\Omega$ contains a
finite number of inhomogeneities, each of the form $z_j + \alpha
B_j$, where $B_j \subset \R^3$ is a bounded, smooth domain
containing the origin.  The total collection of inhomogeneities is
 $ \ds {\cal B}_\alpha
 = \ds \cup_{j=1}^{m} (z_j  + \alpha B_j)$.
  The points $z_j \in \Omega, j=1, \ldots, m,$ which determine the
  location of the inhomogeneities, are assumed to satisfy the
  following inequalities:
\be \label{f1}
 | z_j  - z_l | \geq c_0 > 0, \forall \; j \neq l  \quad
\mbox{ and } \mbox{ dist} (z_j, \partial \Omega) \geq c_0
> 0, \forall \; j,
\ee where $c_0$ is a positive constant.  Assume that $\alpha
>0$, the common order of magnitude of the diameters of the
inhomogeneities, is sufficiently small, that these
 inhomogeneities are disjoint,  and that
their distance to $\R^3 \setminus \overline{\Omega}$ is larger
than $c_0$.\\
Let ${\bf n}(x) \in C^0(\Omega)$ denote the unperturbed refractive
index. We assume that ${\bf n}(x)$ is known on a neighborhood of
the boundary $\partial \Omega$.  Let ${\bf n}_j(x) \in
C^0(\overline{z_j+\alpha B_j})$ denote the refractive index of the
j-th inhomogeneity, $z_j+\alpha B_j$. Introduce the perturbed
potential
\begin{equation}
{\bf n}_\alpha(x)=\left \{ \begin{array}{*{2}{l}}
 {\bf n}(x),\;\;& x \in \Omega \setminus \bar {\cal B}_\alpha,  \\
 {\bf n}_j(x),\;\;& x \in z_j+\alpha B_j, \;j=1 \ldots m.
\end{array}
\right . \label{murhodef}
\end{equation}
Consider the acoustic equation, at fixed frequency, in the
presence of the inhomogeneities ${\cal B}_\alpha$
\[\begin{array}{l}
(\Delta + \omega^2{\bf n}_\alpha) u_\alpha = 0 \mbox{ in } \Omega\\
\nm u_\alpha |_{\partial \Omega} = f \in
\widetilde{H}^{\frac{1}{2}}(\Gamma),
\end{array}
\]
and define the local Dirichlet to Neumann map associated to ${\bf
n}_\alpha$ by :$\Lambda_{{\bf n}_\alpha}(f) = \frac{\partial
u_\alpha}{\partial \nu}|_{\Gamma}$ for all $f \in
\widetilde{H}^{\frac{1}{2}}(\Gamma).$ Where $\widetilde{H}^{\frac{1}{2}}(\Gamma)$ is the trace space\\

Bukhgeim and Uhlmann \cite{BU} used the complex geometrical optics
solutions vanishing on the complementary of $\Gamma$ to prove that
the knowledge of the partial Cauchy data for the Schr\"{o}dinger
equation on any open subset $\Gamma$ of the boundary determines
uniquely the potential $q$ provided that $q\in L^{\infty}(\Omega)$
is known in a neighborhood of the boundary. We may refer to their
work to show our physical version and we use these complex
geometrical optics solutions, constructed in their work, to prove
our
reconstruction procedure.\\

Next, we derive a formula for calculating the refractive index
${\bf n}$ from the partial Dirichlet-to-Neumann map (DN)
associated to ${\bf n}$ and $\Gamma$ by modifying the Nachman's
reconstruction procedure~\cite{Nach}. It turns out that if the
refractive index is a-priori known in a neighborhood of the
boundary the derivation of a reconstruction formula are {\em
surprisingly simple}. Finally, we profit to properties of the
kernel of the introduced operator to explicit more this
reconstruction formula. We consider the inverse problem of
identifying locations of small volume fraction perturbations of
the refractive index and we apply our results to reduce this
inverse problem to calculations of inverse Fourier transforms.
\section{Identification procedure}
Let ${\bf n}_0 \in L^\infty(\Omega)$ be a known function. Assume
that ${\bf n}={\bf n}_0$ almost everywhere in a neighborhood of
$\partial \Omega$. We extend ${\bf n}$ and ${\bf n}_0$ by $0$ in
$\R^3$. Let $\rho \in \R^3\setminus \{0\}$ and define $u_\rho$ to
be the solution of
 \be \begin{array}{l}
 \label{eqr3} \Delta u_\rho =0 \quad \mbox{ in }
\R^3 \setminus \overline{\Omega} , \\
 \nm
 (\frac{1}{\omega^2}\Delta + {\bf n}) u_\rho = 0  \quad \mbox{ in } \Omega ,
 \end{array}
 \ee
we find that $u_\rho |_\Gamma$ solves the (hyper-singular)
integral equation on the open surface $\Gamma$: \be \label{eqr5}
\Lambda_{\bf n}(u_\rho|_\Gamma) + \oint_\Gamma \frac{\partial^2
g^D_{\rho}}{\partial \nu(x) \partial \nu(y)} (x,y)
u_\rho|_\Gamma(y) \; ds(y) = \rho \cdot \nu(x)  e^{x\cdot \rho},
\quad \forall \; x \in \Gamma, \ee where $g^D_{\rho}$ is a well
defined exterior Dirichlet Green's function for $\Delta$.

Now, according to \cite{Alass} and \cite{AR} we can prove the
following global uniqueness result associated to refractive index
in absence of any inhomogeneities (in presence of the
inhomogeneities we may obtain a similar result), as follows.
\begin{prop}\label{prop1}
Suppose that $\alpha=0$. Let ${\bf n}_i$ real-valued in
$L^\infty(\Omega), i=1, 2.$ Assume ${\bf n}_1 ={\bf n}_2$ almost
everywhere in a neighborhood of the boundary $\partial \Omega$ and
$\Lambda_{{\bf n}_1}=\Lambda_{{\bf n}_2}$. Then ${\bf n}_1={\bf
n}_2$
 almost everywhere in $\Omega$.
\end{prop}
\proof Let $-1<\delta<0$ and introduce the weighted $L^2$-space
$$ \ds L^2_\delta(\R^3)= \{ f \in L^2_{loc}(\R^3) | \int_{\R^3} (1
+ |x|^2)^\delta |f(x)|^2\; dx \; < + \infty \}.$$ From \cite{SU},
we know that if $\tilde{{\bf n}}_i$ is an extant function of ${\bf
n}_i$ defined by:
$$
\tilde{{\bf n}}_i(x)=\left \{ \begin{array}{*{2}{l}}
 {\bf n}_i(x),\;\;& x \in \Omega,  \\
 0,\;\;& x \in \R^3\backslash\overline{\Omega},
\end{array}
\right .
$$
then the solutions of $(\frac{1}{\omega^2}\Delta + \tilde{{\bf
n}}_i)v_i=0$ on $\R^3$ can be given by: \be \label{eqv} v_i =
e^{x\cdot \rho_i}(1 + \psi_{{\bf n}_i}(x, \rho_i)), i=1, 2 \ee
 for  $|\rho_i|$ sufficiently large with $\psi_{{\bf n}_i}(\cdot, \rho_i)
 \in L^2_\delta(\R^3)$. Moreover, we have
 \be
 \label{eqv2}
\ds  ||\psi_{{\bf n}_i}(\cdot, \rho_i)||_{L^2_\delta(\R^3)} \leq
 \frac{C}{|\rho_i|}.
 \ee
Now, we set \[ \ds \rho_1 = \frac{\eta}{2} + i (\frac{k
+l}{2})\quad \mbox{ and } \ds \rho_2 =- \frac{\eta}{2} + i
(\frac{k - l}{2})\] where $\eta, k, l \in \R^3$ such that
$\eta\cdot k = \eta \cdot l
= k \cdot l =0,$ and $|\eta|^2 = |k|^2 + |l|^2$.\\
Let $\Omega^\prime \subset \subset \Omega$, $\Omega^\prime$ open
set with $C^2$ boundary and
$\Omega\backslash\overline{\Omega^\prime}$ is connected. Define
$$\widetilde{N}(\Omega) = \Bigr\{v \in H^2(\Omega) \, |\, (\frac{1}{\omega^2}\Delta
+ {\bf n})v = 0 \ \mbox{in} \ \Omega, v= 0 \ \mbox{on} \ \
\Gamma_c \Bigr\}$$ and $$N(\Omega) = \Bigr\{v \in H^2(\Omega) \,
|\, (\frac{1}{\omega^2}\Delta + {\bf n}) v = 0 \ \mbox{in} \
\Omega \Bigr\}.$$ Then, according to \cite{AR} the set
$\widetilde{N}(\Omega)$ is dense, in the $L^2(\Omega^\prime)$
norm, in $N(\Omega)$. On the other hand, by Green's theorem we
have \be \label{eq1} \ds \int_{\Omega^\prime} ({\bf n}_1 -{\bf
n}_2) u_1 u_2 \; dx = \int_\Gamma (\frac{\partial u_1}{\partial
\nu} u_2 - u_1 \frac{\partial u_2}{\partial \nu} )\; ds(x) , \ee
where $ds(x)$ denotes surface measure and $u_1, u_2\in
\widetilde{N}(\Omega)$. Let $z_1 \in
H^1(\Omega)\cap\widetilde{N}(\Omega)$ such that $u_2 |_\Gamma =
z_1 |_\Gamma$. The hypothesis $\Lambda_{{\bf n}_1}=\Lambda_{{\bf
n}_2}$ gives that \be \label{a1} \ds z_1 |_{\Gamma_c} =0, z_1
|_\Gamma = u_2 |_\Gamma \Rightarrow \frac{\partial z_1}{\partial
\nu} |_\Gamma = \frac{\partial u_2}{\partial \nu} |_\Gamma. \ee
Combining relation (\ref{a1}) with (\ref{eq1}), we deduce: \[\ds
\int_{\Omega^\prime} ({\bf n}_1 -{\bf n}_2) u_1 u_2 \; dx =
\int_\Gamma (\frac{\partial u_1}{\partial \nu} u_2 - u_1
\frac{\partial u_2}{\partial \nu} )\; ds \]
\[=\int_\Gamma (\frac{\partial
u_1}{\partial \nu} z_1 - u_1 \frac{\partial z_1}{\partial \nu} )\;
ds. \] By Green's theorem, we have: \[ \ds \int_\Gamma
(\frac{\partial u_1}{\partial \nu} z_1 - u_1 \frac{\partial
z_1}{\partial \nu} )\; ds=\int_{\Omega^\prime} ({\bf n}_1 -{\bf
n}_1) u_1 z_1 \; dx=0,\] which implies
\[\ds\int_{\Omega^\prime} ({\bf n}_1 -{\bf
n}_2) u_1 u_2 \; dx=0.
\]
Now, using density's argument we can approximate any $z_i \in
N(\Omega)$ by elements of $\widetilde{N}(\Omega)$. Therefore the
functions $v_i$ defined in (\ref{eqv}) and belong to $N(\Omega)$
satisfy
\[\ds\int_{\Omega^\prime} ({\bf n}_1 -{\bf
n}_2) z_1 z_2 \; dx=0.
\]
Letting $|l|\rightarrow + \infty$, and using the estimation
(\ref{eqv2}), we deduce:
\[
\ds \widehat{({\bf n}_1 - {\bf n}_2)} (k) = 0 \quad \forall\; k
\in \R^3
\]
which achieves the proof. \square\\ Next, define the double layer
potential
\[
\ds  N_\rho (f) =  \oint_\Gamma \frac{\partial^2
g^D_{\rho}}{\partial \nu(x) \partial \nu(y)} (x,y) f|_\Gamma(y) \;
ds(y), \quad \forall \; f \in \widetilde{H}^{\frac{1}{2}}(\Gamma),
\]
and set: \be\label{rho12}
\begin{array}{l}
\ds \rho_1 = \frac{\eta}{2} + i (\frac{k +l}{2})\\  \ds \rho_2 =-
\frac{\eta}{2} + i (\frac{k - l}{2})
\end{array}
\ee where $\eta, k, l \in \R^3$ such that $\eta\cdot k = \eta
\cdot l = k \cdot l =0,$ and $|\eta|^2 = |k|^2 + |l|^2$. The
following holds.
\begin{lem}\label{lemma1}
Assume that $0$ is not a Dirichlet eigenvalue of
$(\frac{1}{\omega^2}\Delta + {\bf n})$ in $\Omega$ and $\rho_1$ be
given as in (\ref{rho12}). Then, there is a unique solution
$u_{\rho_1} |_\Gamma \in \widetilde{H}^{\frac{1}{2}}(\Gamma)$ of
(\ref{eqr5}) such that \[ \ds u_{\rho_1}|_\Gamma = (\Lambda_{\bf
n} + N_{\rho_1})^{-1} (\rho_1 \cdot \nu(x) e^{x\cdot \rho_1}
|_{\Gamma}).
\]
\end{lem}
Next, we can prove the following reconstruction formula by using
Proposition \ref{prop1}.
\begin{prop} \label{prop2}
Let ${\bf n}_0 \in L^\infty(\Omega)$ be a given function. Assume
that $0$ is not a Dirichlet eigenvalue of
$(\frac{1}{\omega^2}\Delta + {\bf n})$ in $\Omega$ and ${\bf
n}={\bf n}_0$ almost everywhere in a neighborhood of $\partial
\Omega$. Then
\[ \ds \widehat{({\bf n} -{\bf n}_0)}(-k) = \frac{1}{\omega^2}\lim_{|l|\rightarrow + \infty}
 \int_\Gamma   (\Lambda_{\bf n} + N_{\rho_1})^{-1} (\rho_1 \cdot \nu(x) e^{x\cdot \rho_1} |_{\Gamma})
 (\Lambda_{\bf n} - \]
 \[\Lambda_{{\bf n}_0}) (\Lambda_{{\bf n}_0} + N_{\rho_2})^{-1}
 (\rho_2 \cdot \nu(x) e^{x\cdot \rho_2} |_{\Gamma}) \; ds(x),
 \]
where $k$, $\rho_1$ and $\rho_2$ are given as in (\ref{rho12}).
\end{prop}
\proof Let $\rho_i \in \R^3\setminus \{0\}$ ($i=1,2$). Let
$u_{\rho_i} |_\Gamma \in \widetilde{H}^{\frac{1}{2}}(\Gamma)$ be
the solution of (\ref{eqr5}) and define \[ \theta_{{\bf n}}(x,
\rho_i)=e^{-x\cdot \rho_i}u_{\rho_i}(x)-1,
\]
then we have $u_{\rho_i}=e^{-x\cdot \rho_i}(1+\theta_{{\bf n}}(x,
\rho_i))$.\\
We decompose $\rho_s=\tau(\xi+i\eta)$ ($i^2=-1$) with $\xi,\eta\in
\R^3$, $|\xi|=|\eta|=1$. Then we can write:
\[
\Delta_{\rho}\theta_{{\bf n}}={\bf n}\chi(\Omega)\theta_{{\bf
n}}\quad \mbox{ in } \R^3\backslash \Gamma_c,
\]
where $\chi(\Omega)$ is the characteristic function of $\Omega$.
In a small neighborhood $\mathcal{V}(\partial\Omega)$ of the
boundary $\partial\Omega$ let $\nu_0$ denote the normal coordinate
and $\varphi(\frac{\nu_0}{|\rho|})$ a smooth cut-off function
which vanishes on $\mathcal{V}(\partial\Omega)$. The fact that
$\Omega^{\prime}$ is compactly supported in $\Omega$ as mentioned
in the proof of Proposition \ref{prop1}, we have
$\varphi(\frac{\nu_0}{|\rho|})\theta_{{\bf n}}=\theta_{{\bf n}}$
on $\Omega^{\prime}$ for $|\rho|$ sufficiently large. In order to
get the aid of Proposition \ref{prop1} we, firstly, demonstrate
the following estimate \[\ds || \theta_{\bf
n}(\cdot,\rho)||_{L^{2}(\Omega^{\prime})} \leq \frac{C}{|\rho|}.\]
To do this, we may set $\tilde{\theta}_{{\bf
n}}=\varphi(\frac{\nu}{|\rho|})\theta_{{\bf n}}$ in $\R^3$ then
\[
\Delta_{\rho}\tilde{\theta}_{\bf n}={\bf
n}\chi(\Omega)\tilde{\theta}_{\bf n} +\frac{1}{|\rho|}\nabla\cdot
\tilde{\theta}_{\bf n}\cdot\nabla\varphi+ \tilde{\theta}_{\bf
n}\big(\frac{1}{|\rho|^2}+\Delta\varphi+2\frac{\rho}{|\rho|}\cdot\nabla\varphi\big)
\quad \mbox{ in } \R^3.
\]
Obviously that $\tilde{\theta}_{{\bf n}}\in L_{\delta}^{2}$,
therefore by classical results \cite{SU} the following estimate
holds \be\label{x3} \ds || \tilde{\theta}_{\bf
n}||_{L_{\delta}^{2}(\R^3)} \leq \frac{C}{|\tau|},\ee  for some
positive constant $C$ independent of $\tau$. Next, as done in
(\ref{rho12}) we set
\[
\begin{array}{l}
\ds \rho_1 = \frac{\eta}{2} + i (\frac{k +l}{2})\\ \nm \ds \rho_2
=- \frac{\eta}{2} + i (\frac{k - l}{2})
\end{array}
\]
where $\eta, k, l \in \R^3$ such that $\eta\cdot k = \eta \cdot l
= k \cdot l =0,$ and $|\eta|^2 = |k|^2 + |l|^2$.\\
By applying last results, we construct
 \[
 u=u_{\rho}=e^{x\cdot \rho_1}(1+\theta_{{\bf n}}(x,
\rho_1))\quad \mbox{ in } \Omega^{\prime}\] and
 \[ v=v_{\rho}=e^{x\cdot \rho_2}(1+\theta_{{\bf n_0}}(x,
\rho_2))\quad \mbox{ in } \Omega^{\prime},\] to obtain that \be
\label{eqr2} \ds \widehat{({\bf n} -{\bf n}_0)}(-k) =
\lim_{|l|\rightarrow + \infty}
 \int_\Gamma   u_\rho (\Lambda_{\bf n} - \Lambda_{{\bf n}_0}) v_\rho \; ds. \ee
 Finally, by Lemma \ref{lemma1} one can see that the boundary values of the solutions $u_\rho |_\Gamma$ can be recovered from $\Lambda_{\bf
 n}$.The result is then proven. \square\\
The following lemma is useful to give more explanation to our
reconstruction method.
\begin{lem}\label{lem2}
Let $\rho_1$ and $\rho_2$ be given as in (\ref{rho12}), the
following asymptotic behavior holds :
\[N_{\rho_i}=|l|L_i+O(\frac{1}{|l|})\; \mbox{as } |l|\to +\infty,\]
where, for $i\in\{1,2\}$, $L_i$ is a well defined integral
operator on $\widetilde{H}^{\frac{1}{2}}(\Gamma)$.
\end{lem}
\proof We give a brief proof for $i=1$, and the case $i=2$ can be
given similarly. Recall that the kernel of the operator
$N_{\rho_1}$ is $\ds \frac{\partial^2 g^D_{\rho_1}}{\partial
\nu(x) \partial \nu(y)} (x,y)$ where $\ds
g^D_{\rho_1}(x,y)=e^{x\cdot\rho_1}G^D_{\rho_1}(x,y)$ and
$G^D_{\rho_1}(x,y)$ is a solution of the following integral
equation of the first kind : \be\label{eqGD}
-G_{\rho_1}(x)=\int_{\partial\Omega} G_{\rho_1}(x,y)\frac{\partial
G^D_{\rho_1}}{\partial \nu(y)} (x,y)\; ds(y).\ee Here, the
function $G_{\rho_1}$ is given by
\[
G_{\rho_1}(x)=\int_{\R^3}\frac{e^{ix\cdot\xi}}{\xi^2+2i\rho_1\cdot\xi}\;
d\xi.\] On the other hand, one can use (\ref{rho12}) to find that
\[\ds \rho_1\cdot\xi=i\frac{|l|}{2}\Big[ |\xi|\cos(\widehat{\rho_1,\xi})-\frac{(-k\cdot\xi+i\eta\cdot\xi)}{|l|}\Big].\]
Therefore, to continue with the proof we may insert the last
formula into equation (\ref{eqGD}) and we follow the convenable
procedure. \square

Applying Lemma \ref{lem2} to the results found in Proposition
\ref{prop2}, we can prove the following reconstruction formula.
\begin{thm} \label{thm1}
Let ${\bf n}_0 \in L^\infty(\Omega)$ be a given function. Assume
that $0$ is not a Dirichlet eigenvalue of
$(\frac{1}{\omega^2}\Delta + {\bf n})$ in $\Omega$ and ${\bf
n}={\bf n}_0$ almost everywhere in a neighborhood of $\partial
\Omega$. Then
\[ \ds \widehat{({\bf n} -{\bf n}_0)}(-k) = \ds - \frac{\sqrt{2}}{\omega^2}\int_\Gamma   \frac{1}{|\eta|^2}\big(\eta \cdot
\nu(x)|_{\Gamma}\big)^3 e^{ik\cdot x}\; ds(x),
 \]
where $\eta, k\in\R^3$ such that $\eta\cdot k=0$.
\end{thm}

Now, we apply Proposition \ref{prop2} and the asymptotic result
given in \cite{AMV} for identifying efficiently the locations
$\{z_j\}_{j=1}^m$ of the small inhomogeneities ${\cal B}_\alpha$
from the knowledge of the difference between the local DN maps
$\ds \Lambda_{{\bf n}_\alpha} - \Lambda_{\bf n}$ on $\Gamma$. The
following result holds.
\begin{thm}\label{thm2} Suppose that we have (\ref{f1}), let ${\bf n} \in C^0(\Omega)$, ${\bf n}_{\alpha}$ be given by (\ref{murhodef}). Assume
that $0$ is not a Dirichlet eigenvalue of
$(\frac{1}{\omega^2}\Delta + {\bf n}_{\alpha})$ in $\Omega$ and
${\bf n}_{\alpha}={\bf n}$ almost everywhere in a neighborhood of
$\partial \Omega$ (for $\alpha$ sufficiently small). Let $k$, $l$,
$\rho_1$ and $\rho_2$ be as in (\ref{rho12}). Then, the following
identification holds:
\[ \ds \widehat{({\bf n}_{\alpha} -{\bf n})}(-k) = \frac{1}{\omega^2}\lim_{|l|\rightarrow + \infty}
 \int_\Gamma   (\Lambda_{{\bf n}_\alpha} + N_{\rho_1})^{-1} (\rho_1 \cdot \nu(x) e^{x\cdot \rho_1} |_{\Gamma})
 (\Lambda_{{\bf n}_\alpha} -\]
 \[ \Lambda_{{\bf n}}) (\Lambda_{{\bf n}} + N_{\rho_2})^{-1}
 (\rho_2 \cdot \nu(x) e^{x\cdot \rho_2} |_{\Gamma}) \; ds(x)
 =\ds
\frac{\alpha^3}{\omega^2} \sum_{j=1}^m ({\bf n}(z_j) - {\bf
n}_j(z_j)) |B_j| e^{i k \cdot z_j} + o(\alpha^3).
 \]
\end{thm}
As done in Theorem \ref{thm1}, we can deduce the following more
explicit result.
\begin{cor}\label{cor1} Suppose that we have the hypothesis of
Theorem \ref{thm2}. Then, the following identification holds:
\[ \ds - \sqrt{2}\int_\Gamma   \frac{1}{|\eta|^2}\big(\eta \cdot
\nu(x)|_{\Gamma}\big)^3 e^{ik\cdot x}\; ds(x)=\ds \alpha^3
\sum_{j=1}^m ({\bf n}(z_j) - {\bf n}_j(z_j)) |B_j| e^{i k \cdot
z_j} + o(\alpha^3),
 \]
 where $\eta, k\in\R^3$ such that $\eta\cdot k=0$.
\end{cor}

By neglecting the remainders $ o(\alpha^3)$ in Corollary
\ref{cor1}, the locations $\{z_j\}_{j=1}^m$ are obtained as
supports of the inverse Fourier transform of
$$ \ds - \sqrt{2}\int_\Gamma   \frac{1}{|\eta|^2}\big(\eta \cdot
\nu(x)|_{\Gamma}\big)^3 e^{ik\cdot x}\; ds(x).$$ If, we get the
points $\{z_j\}_{j=1}^m$, the values $\{{\bf n}_j(z_j) \}_{j=1}^m$
could be obtained by solving a linear system arising from
Corollary \ref{cor1}.

\vspace{1cm}

{\bf Christian DAVEAU}, -Adresse: {\it D\'epartement de
Math\'ematiques, Site Saint-Martin II,\\BP 222, \& Universit\'e de
Cergy-Pontoise, 95302 Cergy-Pontoise Cedex, France.}\\- Email:
christian.daveau@math.u-cergy.fr\\-Tel : (33) (0)1 34 25 66 72.
-Fax : (33) (0)1 34 25 66 45.\\

{\bf Abdessatar KHELIFI}, -Adresse: {\it D\'epartement de
Math\'ematiques, \& Universit\'e des Sciences de Carthage,
Bizerte, 7021, Tunisie.}\\ -Email: abdessatar.khelifi@fsb.rnu.tn\\
-Tel : (216) 97 53 17 13. -Fax : (216) 72 59 05 66.\\

{\bf Anton SUSHCHENKO}, -Adresse: {\it ETIS \& UMR CNRS 8051, 6
avenue du Ponceau, BP 44, 95014 Cergy-Pontoise Cedex, France }\\
-Email: anton.sushchenko@ensea.fr.

\end{document}